\documentclass[10pt]{article}
\usepackage{graphicx}
\usepackage{amsmath}
\usepackage{amssymb}
\usepackage{caption2}
\usepackage{multirow}
\begin{document}

\begin{center}
{\large {\bf \sc{Two-body strong decays  of the hidden-charm tetraquark molecular states via the QCD sum rules}}} \\[2mm]
Tao Hong, Xiao-Song Yang,  Zhi-Gang Wang\footnote{E-mail: zgwang@aliyun.com.  } \\
Department of Physics, North China Electric Power University, Baoding 071003, P. R. China\\
\end{center}
	
\begin{abstract}	
In this work, we extend our previous work on the $D^*\bar{D}^*$ molecular states with the $J^{PC}=0^{++}$, $1^{+-}$ and $2^{++}$ to investigate their two-body strong decays via the QCD sum rules based on rigorous quark-hadron duality. We obtain the partial decay widths therefore  total widths of the ground  states with the $J^{PC}=0^{++}$, $1^{+-}$ and $2^{++}$, which indicate that it is reasonable to assign  the $X_2(4014)$ as the $D^*\bar{D}^*$ tetraquark molecular states with the $J^{PC}=2^{++}$.
\end{abstract}

\section{Introduction}
In 2003, the Belle collaboration made a ground-breaking observation of the $X(3872)$ in the $\pi^+\pi^-J/\psi$ invariant mass spectrum \cite{Belle:2003nnu}. Since then, extensive experimental and theoretical investigations have been carried out to understand the nature of this exotic state.  In 2013/2015, the LHCb collaboration determined the  quantum numbers of the $X(3872)$ to be $J^{PC}=1^{++}$ in the process $B^+ \to X(3872) K^+$ \cite{LHCb:2013kgk,LHCb:2015jfc}. The $X(3872)$ is located very close to the $D^0 \bar{D}^{*0}$ threshold and has a large decay rate to the $D^0 \bar{D}^{*0}$ pair, which suggests that it might  be a loosely bound molecular  state, therefore its properties are studied extensively  \cite{Mehen:2011ds,Meng:2021kmi,Wu:2021udi,Guo:2014ura,Dong:2009uf,WZG-HT-mole-Zc3900,Kalashnikova:2018vkv,
X3872-mole-Tornqvist-PLB-2004,X3872-mole-Swanson-PLB-2004,
X3872-mole-Mehen-PRD-2007,WZG-mole-IJMPA,
X3872-mole-Grinstein-PRL-2009,X3872-mole-Close-PLB-2004,WZG-mole-Y3940,
X3872-mole-Oset-PRD-2010,
X3872-mole-Wong-PRC-2004,X3872-mole-Petrov-PLB-2006,X3872-mole-Kalashnikova-PRD-2005,
X3872-mole-Braaten-PRD-2010,X3872-mole-Lyubovitskij-PRD-2008,WZG-review}.

In 2021, the Belle collaboration observed weak evidences for two structures in the $\gamma \psi(2S)$ invariant mass distribution of  the two-photon process $\gamma\gamma \to\gamma \psi(2S)$  from the threshold to 4.2 GeV for the first time, one structure was seen  at $3922.4\pm 6.5\pm 2.0\,\rm{MeV}$ with a width of $22\pm 17\pm 4\,\rm{MeV}$, which is consistent with the $X(3915)$ or $\chi_{c2}(3930)$, and the other was seen  at $4014.3\pm 4.0\pm 1.5 \, \rm{MeV}$ with a width of $4\pm 11\pm 6\, \rm{ MeV}$, which is a new  charmonium-like state with a global significance of $2.8\,\sigma$ \cite{X2-4014-Belle-PRD-2022}.

According to the heavy quark spin symmetry,  there maybe exist  an isoscalar  $D^*\bar{D}^*$ molecular state $X(4014)$ with $J^{PC}=2^{++}$ as the partner of the $X(3872)$ \cite{WZG-mole-IJMPA,WZG-mole-Y3940,Nieves:2012tt,Guo:2013sya,Shi:2023mer,HeJ-mole,LiuXH-mole}, which has   a narrow width similar to that of the $X(3872)$ \cite{Guo:2013sya,Shi:2023mer,Cai:2024glz}, and the recent studies on the radiative decays  indicated that the branching  ratio of  $X(4014)\to \gamma \psi(2S)$ to $X(4014)\to\gamma J/ \psi$ is smaller than one, just like that of the $X(3872)$ \cite{Shi:2023mer,KangXW-PLB}. For example, the molecule candidate $X(4014)$ has been studied by several phenomenological models \cite{Nieves:2012tt,Guo:2013sya,Cincioglu:2016fkm,Yu:2024ljg}. It is interesting to view this subject from the QCD directly, and resort to the QCD sum rules to explore the branching fractions of all the $D^*\bar{D}^*$ molecular states in a comprehensive way, thus shed light on the nature of the molecular states.

The QCD sum rules have been successfully applied to study the mass spectrum of the exotic $X$, $Y$ and $Z$ states to determine their natures, whether they are hidden-charm (or hidden-bottom) tetraquark states or hadronic molecular states \cite{WZG-HT-mole-Zc3900,WZG-mole-IJMPA,WZG-review,Wang-tetra-formula,WZG-EPJC-P-2P,WZG-HC-spectrum-PRD,MNielsen-review-1812,Narison-3872,WangHuangtao-2014-PRD, ZhangJR,WangHuang-2014-NPA,Chen-Zhu,C.F.Qiao,WZG-EPJC-P-wave,Azizi-1,Azizi-2,WZG-HC-spectrum-NPB,WZG-XQ-mole-AAPPS,WZG-XQ-mole-EPJA,
WZG-X3872-decay,Wan-BD,Ozdem,WZG-tetra-cc-EPJC,WZG-tetra-cc-APPB,WZG-tetra-psedo-NPB,
Chen-Chen-Zhu,Chen-Chen-Zhu-2,WZG-HB-spectrum-EPJC,Review-QCDSR-Nielsen-PRT-2010,Azizi-Review-2020}.
In our unique scheme focusing on the energy scale dependence of the tetraquark (molecular) states, we  have performed comprehensive studies of the hidden-charm molecular states with the $J^{PC}=0^{++}$, $1^{+-}$, $2^{++}$ \cite{WZG-HT-mole-Zc3900,WZG-mole-IJMPA,WZG-mole-Y3940,WZG-XQ-mole-AAPPS}, doubly-charm molecular (tetraquark)  states with the $J^{P}=0^{+}$, $1^{+}$, $2^{+}$ \cite{WZG-XQ-mole-AAPPS,WZG-XQ-mole-EPJA,WZG-tetra-cc-EPJC,WZG-tetra-cc-APPB},
also hidden-charm tetraquark states with the  $J^{PC}=0^{++}$,  $0^{-+}$, $0^{--}$, $1^{--}$, $1^{-+}$, $1^{+-}$, $2^{++}$ \cite{WZG-EPJC-P-2P,WZG-HC-spectrum-PRD,WZG-EPJC-P-wave,WZG-HC-spectrum-NPB,WZG-tetra-psedo-NPB}, hidden-bottom tetraquark states with the  $J^{PC}=0^{++}$, $1^{+-}$, $2^{++}$ \cite{WZG-HB-spectrum-EPJC}, and made reasonable/suitable  identifications of the existing exotic states. And we observe that the scenario of tetraquark states could accommodate more exotic particles than that of the molecular states.

In our unique scheme of the QCD sum rules, we prefer the pole contributions  about $(40-60)\%$ and the contributions of the vacuum condensates $\leq 1\%$ or $\ll 1\%$ in the Borel windows.  The diquark-antidiquark type tetraquark states $[uc]_A[\bar{d}\bar{c}]_A$ with the $J^{PC}=0^{++}$, $1^{+-}$ and $2^{++}$ have the masses $3.95 \pm 0.09\,\rm{GeV}$, $4.02 \pm 0.09\,\rm{GeV}$ and $4.08 \pm 0.09\,\rm{GeV}$, respectively, where the subscript $A$ denotes the axial-vector diquarks  \cite{WZG-HC-spectrum-PRD}. The central values of the masses have a hierarchy $M_{2^{++}}> M_{1^{+-}}>M_{0^{++}}$, the spin-breaking effects are remarkable.  While the $D^*\bar{D}^*$ type molecular states with the $J^{PC}=0^{++}$, $1^{+-}$ and $2^{++}$ have the masses $4.02 \pm 0.09 \,\rm{GeV}$, $4.02 \pm 0.09\,\rm{GeV}$ and $4.02 \pm 0.09\,\rm{GeV}$, respectively \cite{WZG-mole-IJMPA}. The central values of the masses are almost degenerated, the spin-breaking effects are tiny.   We can assign the  $Z_c(4020)$ as either a tetraquark state or a molecular state with the $J^{PC}=1^{+-}$,  if only the mass is concerned. We should bear in mind, in the QCD sum rules, we choose the local four-quark currents, which couple potentially to the color $\bar{3}3$-type or  $11$-type tetraquark states, they are all compact objects, although we refer to the $11$-type tetraquark states as the molecular states, they are not the usually called  molecular states \cite{WZG-review}.

As the mass alone cannot identify a hadron unambiguously, it can only lead to a crude assessment, we have to deal with the partial decay widths in details. In Ref.\cite{WZG-ZJX-Zc-Decay}, we study the two-body strong decays of the $Z_c(3900)$ as a  tetraquark state with the  $J^{PC}=1^{+-}$ by introducing   rigorous quark-hadron duality in  the three-point QCD sum rules for the first time. Thereafter, the rigorous  duality has been successfully applied to study the strong decays of the exotic states $X(3872)$, $Z_{c}(4020)$, $Z_{cs}(3985/4000)$, $Z_{cs}(4123)$, $Y(4500)$, $X(6552)$, etc in the scenario of tetraquark states \cite{WZG-X3872-decay,WZG-Y4500-decay,WZG-Zcs4123-decay,WZG-Zcs3985-decay,WZG-Y4500-NPB-2024,WZG-YXS-X6552}.
In this work, we extend our previous works to explore the  two-body strong decays of the $D^*\bar{D}^*$ molecular states with the $J^{PC}=0^{++}$, $1^{+-}$ and $2^{++}$ via the QCD sum rules based on the rigorous quark-hadron duality. In calculations,  we take account of both the connected and disconnected Feynman diagrams to ensure accuracy.

The article is organized as follows: in Section 2, we obtain the QCD sum rules for the hadronic coupling constants of the  molecular states; in Section 3, we present the numerical results and  discussions; finally,  we summarize our findings.

\section{ QCD sum rules for the hadronic coupling constants }
Firstly, let us write down the three-point correlation functions,
\begin{eqnarray}\label{CF-0}
\Pi^1(p,q) &=& i^2 \int d^4 x d^4 y e^{ip\cdot x} e^{iq\cdot y} \langle0| T\left\{J^{\eta_c}(x) J^{\pi}(y) J^{0\dagger}(0)\right\}|0\rangle \, , \nonumber\\
\Pi^2_{\alpha}(p,q) &=& i^2 \int d^4 x d^4 y e^{ip\cdot x} e^{iq\cdot y} \langle0| T\left\{J^{\chi_{c1}}_{\alpha}(x) J^{\pi}(y) J^{0\dagger}(0)\right\}|0\rangle  \, , \nonumber\\
\Pi^3_{\alpha\beta}(p,q) &=& i^2 \int d^4 x d^4 y e^{ip\cdot x} e^{iq\cdot y} \langle0| T\left\{J^{J/\psi}_{\alpha}(x) J^{\rho}_{\beta}(y) J^{0\dagger}(0)\right\}|0\rangle \, ,
\end{eqnarray}

\begin{eqnarray}\label{CF-1}
\Pi^4_{\alpha\mu\nu}(p,q) &=& i^2 \int d^4 x d^4 y e^{ip\cdot x} e^{iq\cdot y} \langle0|T\left\{J^{J/\psi}_{\alpha}(x) J^{\pi}(y) J_{\mu\nu}^{1\dagger}(0) \right\}|0\rangle \, , \nonumber\\
\Pi^5_{\beta\mu\nu}(p,q) &=& i^2 \int d^4 x d^4 y e^{ip\cdot x} e^{iq\cdot y} \langle0|T\left\{J^{\eta_c}(x) J^{\rho}_{\beta}(y) J_{\mu\nu}^{1\dagger}(0)\right\}|0\rangle \, , \nonumber\\
\Pi^6_{\alpha\beta\mu\nu}(p,q) &=& i^2 \int d^4 x d^4 y e^{ip\cdot x} e^{iq\cdot y} \langle0|T\left\{J^{h_c}_{\alpha\beta}(x) J^{\pi}(y) J_{\mu\nu}^{1\dagger}(0)\right\}|0\rangle \, ,
\end{eqnarray}

\begin{eqnarray}\label{CF-2}
\Pi^7_{\mu\nu}(p,q) &=& i^2 \int d^4 x d^4 y e^{ip\cdot x} e^{iq\cdot y} \langle0| T\left\{J^{\eta_c}(x) J^{\pi}(y) J_{\mu\nu}^{2\dagger}(0) \right\}|0\rangle \, , \nonumber\\
\Pi^8_{\alpha\mu\nu}(p,q) &=& i^2 \int d^4 x d^4 y e^{ip\cdot x} e^{iq\cdot y} \langle0| T\left\{J^{\chi_{c1}}_{\alpha}(x) J^{\pi}(y) J_{\mu\nu}^{2\dagger}(0)\right\}|0\rangle \, , \nonumber\\
\Pi^9_{\alpha\beta\mu\nu}(p,q) &=& i^2 \int d^4 x d^4 y e^{ip\cdot x} e^{iq\cdot y} \langle0| T\left\{J^{J/\psi}_{\alpha}(x) J^{\rho}_{\beta}(y) J_{\mu\nu}^{2\dagger}(0) \right\}|0\rangle \, ,
\end{eqnarray}
where the currents
\begin{eqnarray}
J^{\eta_c}(x) &=& \bar{c}(x)i\gamma_5 c(x)\, , \nonumber\\
J^{\pi}(x)&=&\bar{u}(x)i\gamma_5  d(x)\, ,\nonumber\\
J^{J/\psi}_{\alpha}(x) &=& \bar{c}(x)\gamma_\alpha c(x)\, , \nonumber\\
J^{\rho}_{\alpha}(x)&=&\bar{u}(x)\gamma_\alpha  d(x)\, ,\nonumber\\
J^{\chi_{c1}}_{\alpha}(x) &=& \bar{c}(x)\gamma_\alpha \gamma_5 c(x)\, , \nonumber\\
J^{h_c}_{\alpha\beta}(x) &=& \bar{c}(x)\sigma_{\alpha\beta} c(x)\, ,
\end{eqnarray}
interpolate the mesons  $\eta_c$, $\pi$, $J/\psi$, $\rho$, $\chi_{c1}$ and $h_c$, respectively,  the currents
\begin{eqnarray}
J^0(x) &=& \bar{u}(x)\gamma_\mu c(x)\, \bar{c}(x)\gamma^\mu  d(x) \, ,\nonumber \\
J_{\mu\nu}^1(x) &=&\frac{1}{\sqrt{2}}\Big[\bar{u}(x)\gamma_\mu c(x)\, \bar{c}(x)\gamma_\nu d(x)-\bar{u}(x)\gamma_\nu c(x)\, \bar{c}(x) \gamma_\mu d(x) \Big] \, ,\nonumber\\
J_{\mu\nu}^2(x) &=&\frac{1}{\sqrt{2}}\Big[\bar{u}(x)\gamma_\mu c(x)\, \bar{c}(x)\gamma_\nu d(x)+\bar{u}(x)\gamma_\nu c(x)\, \bar{c}(x) \gamma_\mu d(x) \Big] \, ,
\end{eqnarray}
interpolate the hidden-charm molecular states with the spins $J=0$, $1$ and $2$, respectively. We take those correlation functions to study the hadronic coupling constants $G_{X_0\eta_c \pi}$, $G_{X_0\chi_{c1} \pi}$, $G_{X_0J/\psi  \rho}$, $G_{X_1J/\psi \pi}$, $G_{X_1\eta_c \rho}$, $G_{X_1h_c \pi}$, $G_{X_2\eta_c \pi}$, $G_{X_2\chi_{c1} \pi}$, and $G_{X_2J/\psi  \rho}$, respectively. In this work, we take the isospin limit for simplicity.  

We insert a complete set of intermediate hadronic states with the same quantum numbers as the currents into Eqs.\eqref{CF-0}-\eqref{CF-2} to analyze the correlation functions at the hadron side, then we explicitly isolate the contributions of the ground states \cite{SVZ79,Reinders85},	
\begin{eqnarray}\label{Hadron-CT-1}
\Pi^1(p,q)&=& \frac{\lambda_{X_0} f_{\eta_c} m_{\eta_c}^2  f_{\pi} m_{\pi}^2 G_{X_0\eta_c \pi} }{2m_c(m_u+m_d)(m_{X_0}^2-p^{\prime2})(m_{\eta_c}^2-p^2)(m_{\pi}^2-q^2)}  + \cdots\, ,\nonumber\\
&=&\Pi_{1}(p^{\prime2},p^2,q^2)    + \cdots\, ,
\end{eqnarray}
			
\begin{eqnarray}\label{Hadron-CT-2}
\Pi^2_{\alpha}(p,q)&=& \frac{\lambda_{X_0} f_{\chi_{c1}} m_{\chi_{c1}} f_{\pi} m_{\pi}^2G_{X_0\chi_{c1}\pi} }{(m_u+m_d)(m_{X_0}^2-p^{\prime2})(m_{\chi_{c1}}^2-p^2)(m_{\pi}^2-q^2)}\,iq_\alpha + \cdots\, , \nonumber\\
&=&\Pi_{2}(p^{\prime2},p^2,q^2)\, (iq_\alpha)    + \cdots\, ,
\end{eqnarray}
		
\begin{eqnarray}\label{Hadron-CT-3}
\Pi^3_{\alpha\beta}(p,q)&=& \frac{\lambda_{X_0} f_{J/\psi} m_{J/\psi} f_{\rho}m_{\rho} G_{X_0J/\psi \rho}   }{(m_{X_0}^2-p^{\prime2})(m_{J/\psi}^2-p^2)(m_{\rho}^2-q^2)}  g_{\alpha\beta}    + \cdots\, , \nonumber\\		
&=&\Pi_{3}(p^{\prime2},p^2,q^2)\,   g_{\alpha\beta}  + \cdots\, ,
\end{eqnarray}
	
\begin{eqnarray}
P_{A}^{\mu\nu\alpha^\prime\beta^\prime}(p^\prime)\,\epsilon_{\alpha^\prime\beta^\prime}{}^{\alpha\tau}\,p_\tau\,\Pi^4_{\alpha\mu\nu}(p,q)&=&
		\widetilde{\Pi}_4(p^{\prime2},p^2,q^2)\left(p^2+p\cdot q\right)\, ,
\end{eqnarray}
		
\begin{eqnarray}\label{Hadron-CT-4}
\Pi_4(p^{\prime2},p^2,q^2)&=&\widetilde{\Pi}_4(p^{\prime2},p^2,q^2)\,p\cdot q,\nonumber\\
		&=&\frac{\lambda_{X_1} f_{J/\psi}m_{J/\psi}f_{\pi} m_{\pi}^2 G_{X_1J/\psi \pi} }{m_{X_1}(m_u+m_d)(m_{X_1}^2-p^{\prime2})(m_{J/\psi}^2-p^2)(m_{\pi}^2-q^2)}\,p\cdot q   + \cdots\ ,
\end{eqnarray}
	
\begin{eqnarray}
 P_{A}^{\mu\nu\alpha^\prime\beta^\prime}(p^\prime)\,\epsilon_{\alpha^\prime\beta^\prime}{}^{\beta\tau}\,q_\tau\,\Pi^5_{\beta\mu\nu}(p,q)&=&
	\widetilde{\Pi}_5(p^{\prime2},p^2,q^2)\left(p\cdot q+q^2\right)\, ,
\end{eqnarray}
	
\begin{eqnarray}\label{Hadron-CT-5}
	\Pi_5(p^{\prime2},p^2,q^2)&=&\widetilde{\Pi}_5(p^{\prime2},p^2,q^2)\,p\cdot q,\nonumber\\
	&=&\frac{\lambda_{X_1} f_{\eta_c}m_{\eta_c}^2f_{\rho} m_{\rho} G_{X_1\eta_c \rho} }{2m_cm_{X_1}(m_{X_1}^2-p^{\prime2})(m_{\eta_c}^2-p^2)(m_{\rho}^2-q^2)}\,p\cdot q + \cdots\, ,
\end{eqnarray}
	
\begin{eqnarray}
P_{A}^{\mu\nu\mu^\prime\nu^\prime}(p)P_{A}^{\alpha\beta\alpha^\prime\beta^\prime}(p^\prime)\,\epsilon_{\mu^\prime\nu^\prime\alpha^\prime\beta^\prime}\,
	 \Pi^6_{\alpha\beta\mu\nu}(p,q)&=&i\widetilde{\Pi}_6(p^{\prime2},p^2,q^2)\left(p\cdot q^2-p^2q^2\right)\, ,
\end{eqnarray}
	
\begin{eqnarray}\label{Hadron-CT-6}
\Pi_6(p^{\prime2},p^2,q^2)&=&\widetilde{\Pi}_6(p^{\prime2},p^2,q^2)\,ip \cdot q^2\, ,\nonumber\\
	&=&\frac{2\lambda_{X_1} f_{h_c}f_{\pi} m_{\pi}^2G_{X_1h_c \pi} }{9m_{X_1}(m_u+m_d)(m_{X_1}^2-p^{\prime2})(m_{h_c}^2-p^2)(m_{\pi}^2-q^2)}\,ip \cdot q^2+ \cdots\, ,
\end{eqnarray}
	
\begin{eqnarray}\label{Hadron-CT-7}
\Pi^7_{\mu\nu}(p,q)&=& -\frac{\lambda_{X_2}f_{\eta_c} m_{\eta_c}^2f_{\pi} m_{\pi}^2 (m_{X_2}^2-m_{\eta_c}^2) G_{X_2\eta_c \pi} }{2m_c m_{X_2}^2 (m_u+m_d)(m_{X_2}^2-p^{\prime2})(m_{\eta_c}^2-p^2)(m_{\pi}^2-q^2)}\,p_\mu q_\nu +  \cdots\, , \nonumber\\
	&=&\Pi_{7}(p^{\prime2},p^2,q^2)\left(-p_{\mu}q_{\nu}\right)  + \cdots\, ,
\end{eqnarray}

\begin{eqnarray}\label{Hadron-CT-8}
\Pi^8_{\alpha\mu\nu}(p,q)&=& \frac{\lambda_{X_2} f_{\chi_{c1}} m_{\chi_{c1}}  f_{\pi} m_{\pi}^2 G_{X_2\chi_{c1}\pi} }{(m_u+m_d)(m_{X_2}^2-p^{\prime2})(m_{\chi_{c1}}^2-p^2)(m_{\pi}^2-q^2)}\left(\frac{q_\alpha g_{\mu\nu}}{3}-\frac{q_\mu g_{\alpha\nu}}{2}-\frac{q_\nu g_{\alpha\mu}}{2}\right) + \cdots\, , \nonumber\\
	&=&\Pi_{8}(p^{\prime2},p^2,q^2)\left(\frac{q_\alpha g_{\mu\nu}}{3}-\frac{q_\mu g_{\alpha\nu}}{2}-\frac{q_\nu g_{\alpha\mu}}{2}\right) + \cdots\, ,
\end{eqnarray}
	
\begin{eqnarray}\label{Hadron-CT-9}
\Pi^9_{\alpha\beta\mu\nu}(p,q)&=& -\frac{\lambda_{X_2} f_{J/\psi} m_{J/\psi}  f_{\rho} m_{\rho} G_{X_2J/\psi\rho} }{2(m_{X_2}^2-p^{\prime2})(m_{J/\psi}^2-p^2)(m_{\rho}^2-q^2)}\left( g_{\alpha\mu}g_{\beta\nu}+g_{\alpha\nu}g_{\beta\mu}\right)  + \cdots\, , \nonumber \\
	&=&\Pi_{9}(p^{\prime2},p^2,q^2)\left( -g_{\alpha\mu}g_{\beta\nu}-g_{\alpha\nu}g_{\beta\mu}\right)  + \cdots\, ,
\end{eqnarray}		
where
\begin{eqnarray}
P_{A}^{\mu\nu\alpha\beta}(p)&=&\frac{1}{6}\left( g^{\mu\alpha}-\frac{p^\mu p^\alpha}{p^2}\right)\left( g^{\nu\beta}-\frac{p^\nu p^\beta}{p^2}\right)\, ,
\end{eqnarray}
the decay constants or pole residues are defined by,
\begin{eqnarray}
\langle0|J^{\eta_c}(0)|\eta_c(p)\rangle&=&\frac{f_{\eta_c} m_{\eta_c}^2}{2m_c}  \,\, , \nonumber \\	
\langle0|J^{\pi}(0)|\pi(p)\rangle&=&\frac{f_{\pi} m_{\pi}^2}{m_u+m_d}  \,\, , \nonumber \\
\langle0|J_{\mu}^{J/\psi}(0)|J/\psi(p)\rangle&=&f_{J/\psi} m_{J/\psi} \,\xi_\mu \,\, , \nonumber \\
\langle0|J_{\nu}^{\rho}(0)|\rho(p)\rangle&=&f_{\rho} m_{\rho} \,\zeta_\nu \,\, , \nonumber \\
\langle0|J_{\mu\nu}^{h_c}(0)|h_c(p)\rangle&=&f_{h_c} \epsilon_{\mu\nu\alpha\beta}\, p^\alpha \xi^\beta \,\, , \nonumber \\
\langle0|J_{\mu}^{\chi_c}(0)|\chi_c(p)\rangle&=&f_{\chi_c} m_{\chi_c}\, \zeta_\mu \,\, ,
\end{eqnarray}
\begin{eqnarray}
\langle 0|J^0(0)|X_0 (p)\rangle &=& \lambda_{X_0}     \, , \nonumber\\
\langle 0|J^1_{\mu\nu}(0)|X_1(p)\rangle &=& \tilde{\lambda}_{X_1} \, \epsilon_{\mu\nu\alpha\beta} \, \varepsilon^{\alpha}p^{\beta}\, , \nonumber\\
\langle 0|J^2_{\mu\nu}(0)|X_2 (p)\rangle &=& \lambda_{X_2} \, \varepsilon_{\mu\nu}   \, ,
\end{eqnarray}
$\tilde{\lambda}_{X_1}m_{X_1}=\lambda_{X_1}$, the hadronic coupling constants are defined by,
\begin{eqnarray}
\langle \eta_c(p)\pi(q)|X_0(p)\rangle&=&  i G_{X_0\eta_c\pi}\, ,\nonumber \\
\langle J/\psi(p)\rho(q)|X_0(p)\rangle&=& i  \xi^* \cdot \xi^* \, G_{X_0J/\psi \rho}\, ,\nonumber \\
\langle \chi_c(p)\pi(q)|X_0(p)\rangle&=&  \zeta^* \cdot q \,G_{X_0\chi_c \pi}\, ,
\end{eqnarray}
\begin{eqnarray}
\langle \eta_c(p)\rho(q)|X_1(p)\rangle&=& i\xi^* \cdot \varepsilon \,G_{X_1\eta_c \rho}\, ,\nonumber \\
\langle J/\psi(p)\pi(q)|X_1(p)\rangle&=& i\xi^* \cdot \varepsilon \,G_{X_1J/\psi\pi}\, ,\nonumber \\
\langle h_c(p)\pi(q)|X_1(p)\rangle&=& \epsilon^{\lambda\tau\rho\sigma}p_{\lambda}\xi^*_{\tau}p^\prime_\rho\varepsilon_\sigma  \,G_{X_1h_c \pi}\, ,
\end{eqnarray}
\begin{eqnarray}
\langle \eta_c(p)\pi(q)|X_2(p)\rangle&=& -i \varepsilon_{\mu\nu}p^{\mu}q^{\nu} \,G_{X_2\eta_c\pi}\, ,\nonumber \\
		\langle \chi_c(p)\pi(q)|X_2(p)\rangle&=& -i \varepsilon_{\alpha\beta} \xi^{*\alpha} q^{\beta}\,G_{X_2\chi_c\pi}\, ,\nonumber \\
		\langle J/\psi(p)\rho(q)|X_2(p)\rangle&=& -i \varepsilon^{\alpha\beta} \xi^*_\alpha  \xi^*_\beta \, G_{X_2J/\psi \rho}\, ,
\end{eqnarray}
the $\varepsilon_{\alpha}$ and $\varepsilon_{\mu\nu} $ represent the polarization vectors of the tetraquark molecular states with the spins $J=1$ and $2$, respectively. The $\xi_\mu$ and $\zeta_\nu$ represent the polarization vectors of the traditional  mesons with the spin $J=1$. Since the currents $J_{\alpha\beta}^{h_c}(x)$ and $J^1_{\mu\nu}(x)$   couple potentially to charmonia/tetraquarks with both the $J^{PC}=1^{+-}$ and $1^{--}$, we introduce the projector $P_{A}^{\mu\nu\alpha\beta}(p)$ to extract the states with the $J^{PC}=1^{+-}$ \cite{WZG-HC-spectrum-PRD,WZG-HC-spectrum-NPB,WZG-HB-spectrum-EPJC}. In this work, we investigate the hadronic coupling constants using the components  $\Pi_{i}(p^{\prime2},p^2,q^2)$ with $i=1$, $\cdots$, $9$, and try to avoid any contamination.
	
It is the time to obtain the hadronic  spectral densities $\rho_H(s^\prime,s,u)$ through triple  dispersion relation,
\begin{eqnarray}\label{dispersion-3}
		\Pi_{H}(p^{\prime2},p^2,q^2)&=&\int_{4m_c^2}^\infty ds^{\prime} \int_{4m_c^2}^\infty ds \int_{0}^\infty du \frac{\rho_{H}(s^\prime,s,u)}{(s^\prime-p^{\prime2})(s-p^2)(u-q^2)}\, ,
\end{eqnarray}
where
\begin{eqnarray}
\rho_{H}(s^\prime,s,u)&=&{\lim_{\epsilon_3\to 0}}\,\,{\lim_{\epsilon_2\to 0}} \,\,{\lim_{\epsilon_1\to 0}}\,\,\frac{ {\rm Im}_{s^\prime}\, {\rm Im}_{s}\,{\rm Im}_{u}\,\Pi_{H}(s^\prime+i\epsilon_3,s+i\epsilon_2,u+i\epsilon_1) }{\pi^3} \, ,
\end{eqnarray}
we add the subscript $H$ to stand for  the components $\Pi_{i}(p^{\prime2},p^2,q^2)$ with $i=1-9$ at the hadron side. Although the variables $p^\prime$, $p$ and $q$ have the relation $p^\prime=p+q$, it is feasible to take the $p^{\prime2}$, $p^2$ and $q^2$ as free variables to determine the spectral densities, and we can obtain a nonzero  imaginary part indeed for all the variables $p^{\prime2}$, $p^2$ and $q^2$.
		
At the QCD side, we contract all the quark fields with the Wick's theorem and take account of the perturbative terms, quark condensate, gluon condensate  and quark-gluon mixed condensate contributions, and obtain the QCD spectral densities of the components $\Pi_{i}(p^{\prime2},p^2,q^2)$ through double dispersion relation,
\begin{eqnarray}\label{dispersion-2}
\Pi_{QCD}(p^{\prime2},p^2,q^2)&=& \int_{4m_c^2}^\infty ds \int_{0}^\infty du \frac{\rho_{QCD}(p^{\prime2},s,u)}{(s-p^2)(u-q^2)}\, ,
\end{eqnarray}
as
\begin{eqnarray}
{\rm lim}_{\epsilon_3 \to 0}{\rm Im}_{s^\prime}\,\Pi_{QCD}(s^\prime+i\epsilon_3,p^2,q^2)&=&0\, .
\end{eqnarray}
		
At the hadron side, there is a triple dispersion relation, see Eq.\eqref{dispersion-3}, while at the QCD side, there is only a double dispersion relation, see Eq.\eqref{dispersion-2}. These relations do not match with each other channel by channel without some tricks. To achieve this goal, we firstly integrate over $ds^\prime$ at the hadron side, and then match the hadron side with the QCD side of the components $\Pi_{i}(p^{\prime2},p^2,q^2)$ below the continuum thresholds $s_0$ and $u_0$ respectively  to establish quark-hadron duality rigorously \cite{WZG-ZJX-Zc-Decay,WZG-Y4660-Decay},
\begin{eqnarray}\label{Duality}
\int_{4m_c^2}^{s_0}ds \int_{0}^{u_0}du  \left[ \int_{4m_c^2}^{\infty}ds^\prime  \frac{\rho_H(s^\prime,s,u)}{(s^\prime-p^{\prime2})(s-p^2)(u-q^2)} \right] &=&\int_{4m_c^2}^{s_{0}}ds \int_{0}^{u_0}du  \frac{\rho_{QCD}(s,u)}{(s-p^2)(u-q^2)}
		\, . \nonumber\\
\end{eqnarray}
For clearness,  we write down the hadron representation explicitly,
\begin{eqnarray}\label{HS-CT-1}
\Pi_{1}(p^{\prime2},p^2,q^2)&=& \frac{\lambda_{X_0} f_{\eta_c} m_{\eta_c}^2  f_{\pi} m_{\pi}^2 G_{X_0\eta_c \pi} }{2m_c(m_u+m_d)(m_{X_0}^2-p^{\prime2})(m_{\eta_c}^2-p^2)(m_{\pi}^2-q^2)}
		+\frac{C_1 }{(m_{\eta_c}^2-p^2)(m_{\pi}^2-q^2)}
 \, ,\nonumber\\
\end{eqnarray}
		
\begin{eqnarray}\label{HS-CT-2}
\Pi_{2}(p^{\prime2},p^2,q^2)&=& \frac{\lambda_{X_0} f_{\chi_{c1}} m_{\chi_{c1}} f_{\pi} m_{\pi}^2G_{X_0\chi_{c1}\pi} }{(m_u+m_d)(m_{X_0}^2-p^{\prime2})(m_{\chi_{c1}}^2-p^2)(m_{\pi}^2-q^2)}
			+\frac{C_2 }{(m_{\chi_{c1}}^2-p^2)(m_{\pi}^2-q^2)}
			 \, ,\nonumber\\
\end{eqnarray}
\begin{eqnarray}\label{HS-CT-3}
\Pi_{3}(p^{\prime2},p^2,q^2)&=& \frac{\lambda_{X_0} f_{J/\psi} m_{J/\psi} f_{\rho}m_{\rho} G_{X_0J/\psi \rho}   }{(m_{X_0}^2-p^{\prime2})(m_{J/\psi}^2-p^2)(m_{\rho}^2-q^2)}
				+\frac{C_3 }{(m_{J/\psi}^2-p^2)(m_{\rho}^2-q^2)} \, ,\nonumber\\
\end{eqnarray}
\begin{eqnarray}\label{HS-CT-4}
\Pi_{4}(p^{\prime2},p^2,q^2)&=&\frac{\lambda_{X_1} f_{J/\psi}m_{J/\psi}f_{\pi} m_{\pi}^2 G_{X_1J/\psi \pi} }{m_{X_1}(m_u+m_d)(m_{X_1}^2-p^{\prime2})(m_{J/\psi}^2-p^2)(m_{\pi}^2-q^2)}+\frac{C_4 }{(m_{J/\psi}^2-p^2)(m_{\pi}^2-q^2)}  \, ,\nonumber\\
\end{eqnarray}
\begin{eqnarray}\label{HS-CT-5}
\Pi_{5}(p^{\prime2},p^2,q^2)&=&\frac{\lambda_{X_1} f_{\eta_c}m_{\eta_c}^2f_{\rho} m_{\rho} G_{X_1 \eta_c \rho} }{2m_cm_{X_1}(m_{X_1}^2-p^{\prime2})(m_{\eta_c}^2-p^2)(m_{\rho}^2-q^2)}+\frac{C_5 }{(m_{\eta_c}^2-p^2)(m_{\rho}^2-q^2)}  \, ,\nonumber\\
\end{eqnarray}
\begin{eqnarray}\label{HS-CT-6}
\Pi_{6}(p^{\prime2},p^2,q^2)&=&\frac{2 \lambda_{X_1} f_{h_c}f_{\pi} m_{\pi}^2G_{X_1h_c \pi} }{9m_{X_1}(m_u+m_d)(m_{X_1}^2-p^{\prime2})(m_{h_c}^2-p^2)(m_{\pi}^2-q^2)}+\frac{C_6 }{(m_{h_c}^2-p^2)(m_{\pi}^2-q^2)}  \, ,\nonumber\\
\end{eqnarray}
\begin{eqnarray}\label{HS-CT-7}
\Pi_{7}(p^{\prime2},p^2,q^2)&=&\frac{\lambda_{X_2}f_{\eta_c} m_{\eta_c}^2f_{\pi} m_{\pi}^2 (m_{X_2}^2-m_{\eta_c}^2) G_{X_2\eta_c \pi} }{2m_c m_{X_2}^2(m_u+m_d)(m_{X_2}^2-p^{\prime2})(m_{\eta_c}^2-p^2)(m_{\pi}^2-q^2)}+\frac{C_7 }{(m_{\eta_c}^2-p^2)(m_{\pi}^2-q^2)}  \, ,\nonumber\\
\end{eqnarray}
\begin{eqnarray}\label{HS-CT-8}
\Pi_{8}(p^{\prime2},p^2,q^2)&=& \frac{\lambda_{X_2} f_{\chi_{c1}} m_{\chi_{c1}}  f_{\pi} m_{\pi}^2 G_{X_2\chi_{c1}\pi} }{(m_u+m_d)(m_{X_2}^2-p^{\prime2})(m_{\chi_{c1}}^2-p^2)(m_{\pi}^2-q^2)} +\frac{C_8 }{(m_{\chi_{c1}}^2-p^2)(m_{\pi}^2-q^2)}  \, ,\nonumber\\
\end{eqnarray}
\begin{eqnarray}\label{HS-CT-9}
\Pi_{9}(p^{\prime2},p^2,q^2)&=&\frac{\lambda_{X_2} f_{J/\psi} m_{J/\psi}  f_{\rho} m_{\rho} G_{X_2J/\psi\rho} }{2(m_{X_2}^2-p^{\prime2})(m_{J/\psi}^2-p^2)(m_{\rho}^2-q^2)}+\frac{C_9 }{(m_{J/\psi}^2-p^2)(m_{\rho}^2-q^2)}  \, ,\nonumber\\
\end{eqnarray}
where we introduce the parameters $C_{i}$ with $i=1-9$ to stand for  all the contributions involving   the higher resonances and continuum states in the $s^\prime$ channel.
				
We set $p^{\prime2}=p^2$ in the components $\Pi_H(p^{\prime 2},p^2,q^2)$, and perform double Borel transformation  in regard  to the variables  $P^2=-p^2$ and $Q^2=-q^2$, respectively. The spectral densities $\rho_{H}(s^\prime,s,u)$ and $\rho_{QCD}(s,u)$ are physical, while the variables $p^{\prime2}$, $p^2$ and $q^2$ in Eq.\eqref{Duality} are free parameters, as we perform the operator product expansion at the large space-like regions $-p^2 \to \infty$ and $-q^2 \to \infty$. Generally speaking, we can set $p^{\prime2}=\alpha p^2$ or $\alpha q^2$ with $\alpha$ to be a finite quantity. Considering the mass poles at $s^\prime=m^2_{X_{0,1,2}}$, $s=m^2_{\eta_c,J/\psi,\chi_{c1},h_c}$ and $u=m^2_{\pi,\rho}$ have the relations $s^\prime=s$ approximately, we can set $\alpha=1$.
		
Then we set $T_1^2=T_2^2=T^2$  to obtain nine QCD sum rules,
\begin{eqnarray}\label{QCDSR-X0eta-pi}	
&&\frac{\lambda_{X_0\eta_c \pi} G_{X_0\eta_c\pi}}{ {m}^2_{X_0}-m^2_{\eta_c}}\left[\exp\left(-\frac{m^2_{\eta_c}}{T^2}\right) -\exp\left(-\frac{{m}^2_{X_0}}{T^2}\right) \right] \exp\left(-\frac{m^2_{\pi}}{T^2}\right)+C_{1} \exp\left(-\frac{m^2_{\eta_c}}{T^2}-\frac{m^2_{\pi}}{T^2}\right)  \nonumber\\
&=&-\frac{3}{64\pi^4} \int^{s_{\eta_c}^0}_{4m^2_c} ds \int^{s_{\pi}^0}_{0} du su \sqrt{1-\frac{4m^2_c}{s}} \exp\left(-\frac{s+u}{T^2}\right) \nonumber\\
	&&-\frac{ m^4_c }{8\pi^2}\langle\frac{\alpha_{s}GG}{\pi}\rangle \int^{s_{\eta_c}^0}_{4m^2_c} ds \int^{s_{\pi}^0}_{0} du\frac{su \left(2m^2_c-s\right)} { \sqrt{s\left(s-4m^2_c\right)}^5} \exp\left(-\frac{s+u}{T^2}\right) \nonumber\\
	&&-\frac{ m^2_c }{16\pi^2}\langle\frac{\alpha_{s}GG}{\pi}\rangle \int^{s_{\eta_c}^0}_{4m^2_c} ds \int^{s_{\pi}^0}_{0} du\frac{u \left(m^2_c-s\right)} { \sqrt{s\left(s-4m^2_c\right)}^3} \exp\left(-\frac{s+u}{T^2}\right) \, ,		
\end{eqnarray}
where we introduce the notations,	
\begin{eqnarray}
\lambda_{X_0\eta_c\pi}&=\frac{\lambda_{X_0} f_{\eta_c} m^2_{\eta_c}f_{\pi} m^2_{\pi}}{2m_c(m_u+m_d)}\, ,
\end{eqnarray}
and the other eight QCD sum rules are given explicitly in the Appendix. There exists an unknown parameter $C_i$ in each QCD sum rule, we take the $C_{i}$ as free parameters and adjust the suitable values to obtain flat Borel platforms for the hadronic coupling constants
 \cite{WZG-ZJX-Zc-Decay,WZG-Y4500-decay,WZG-Zcs4123-decay,WZG-Zcs3985-decay,WZG-Y4660-Decay}. In calculations, we observe that there exist endpoint divergences at the thresholds  $s=4m_c^2$  due to powers of $s-4m_c^2$ in the denominators, we make the replacements $s-4m_c^2\to s-4m_c^2+\Delta^2$ with $\Delta^2=m_c^2$ to regulate the divergences \cite{WZG-6c-IJMPA,Wang:2021xao}. As the gluon condensates and quark-gluon mixed condensates make tiny contributions, such regulations work well.
	
\section{Numerical results and discussions}
At the QCD side, we take the standard gluon condensate $\langle \frac{\alpha_sGG}{\pi}\rangle=0.012\pm0.004\,\rm{GeV}^4$ \cite{SVZ79,Reinders85,Colangelo-Review}
and take the $\overline{MS}$ mass $m_{c}(m_c)=(1.275\pm0.025)\,\rm{GeV}$  from the Particle Data Group \cite{PDG}. And we take the conventional vacuum condensates $\langle \bar{q}q \rangle=-(0.24\pm 0.01\, \rm{GeV})^3$,
$\langle\bar{q}g_s\sigma G q \rangle=m_0^2\langle \bar{q}q \rangle$,
$m_0^2=(0.8 \pm 0.1)\,\rm{GeV}^2$ at the energy scale $\mu=1\, \rm{GeV}$ \cite{SVZ79,Reinders85,Colangelo-Review}. We set $m_u=m_d=0$ and take account of the energy-scale dependence from re-normalization group equation,
\begin{eqnarray}
	\langle\bar{q}q \rangle(\mu)&=&\langle\bar{q}q \rangle({\rm 1GeV})\left[\frac{\alpha_{s}({\rm 1GeV})}{\alpha_{s}(\mu)}\right]^{\frac{12}{33-2n_f}}\, , \nonumber\\
	\langle\bar{q}g_s \sigma Gq \rangle(\mu)&=&\langle\bar{q}g_s \sigma Gq \rangle({\rm 1GeV})\left[\frac{\alpha_{s}({\rm 1GeV})}{\alpha_{s}(\mu)}\right]^{\frac{2}{33-2n_f}}\, , \nonumber\\	
	 m_c(\mu)&=&m_c(m_c)\left[\frac{\alpha_{s}(\mu)}{\alpha_{s}(m_c)}\right]^{\frac{12}{33-2n_f}} \, ,\nonumber\\
	\alpha_s(\mu)&=&\frac{1}{b_0t}\left[1-\frac{b_1}{b_0^2}\frac{\log t}{t} +\frac{b_1^2(\log^2{t}-\log{t}-1)+b_0b_2}{b_0^4t^2}\right]\, ,
\end{eqnarray}
where $t=\log \frac{\mu^2}{\Lambda^2}$, $b_0=\frac{33-2n_f}{12\pi}$, $b_1=\frac{153-19n_f}{24\pi^2}$, $b_2=\frac{2857-\frac{5033}{9}n_f+\frac{325}{27}n_f^2}{128\pi^3}$,  $\Lambda=213\,\rm{MeV}$, $296\,\rm{MeV}$  and  $339\,\rm{MeV}$ for the flavors  $n_f=5$, $4$ and $3$, respectively \cite{PDG,Narison-mix}.
As we study the hidden-charm tetraquarks, so we choose the flavor numbers $n_f=4$.

At the hadron side, we take $m_{\eta_c}=2.9834\,\rm{GeV}$, $m_{\pi^0}=0.13498\,\rm{GeV}$, $m_{J/\psi}=3.0969\,\rm{GeV}$, $m_{\rho}=0.77526\,\rm{GeV}$, $m_{h_c}=3.525\,\rm{GeV}$, $m_{\chi_{c1}}=3.51067\,\rm{GeV}$ from the Particle Data Group \cite{PDG}. The QCD sum rules allow us to  reproduce the experimental masses of the ground state conventional mesons,  for simplicity, we adopt the precise masses from  the Particle Data Group. We take the values  $s^0_{\pi}=(0.85\,\rm{GeV})^2$, $s^0_{\rho}=(1.2\,\rm{GeV})^2$, $s^0_{h_c}=(3.9\,\rm{GeV})^2$, $s^0_{\chi_{c1}}=(3.9\,\rm{GeV})^2$, $s^0_{J/\psi}=(3.6\,\rm{GeV})^2$, $s^0_{\eta_c}=(3.5\,\rm{GeV})^2$, $f_{\pi}=0.130\,\rm{GeV}$ \cite{Colangelo-Review,PDG},
$f_{h_c}=0.235\,\rm{GeV}$, $f_{J/\psi}=0.418 \,\rm{GeV}$, $f_{\eta_c}=0.387\,\rm{GeV}$ \cite{Becirevic},
$f_{\rho}=0.215\,\rm{GeV} $ \cite{PBall-decay-Kv},
$f_{\chi_{c1}}=0.338\,\rm{GeV}$ \cite{Charmonium-PRT}, $M_{X_0}=4.02\,\rm{GeV}$, $\lambda_{X_0}=4.30\times 10^{-1}\,\rm{GeV}^5$, $M_{X_1}=4.02\,\rm{GeV}$, $\lambda_{X_1}=2.33\times 10^{-1}\,\rm{GeV}^5$, $M_{X_2}=4.02\,\rm{GeV}$, $\lambda_{X_2}=3.29\times 10^{-1}\,\rm{GeV}^5$ from the QCD sum rules \cite{WZG-mole-IJMPA}, and  $f_{\pi}m^2_{\pi}/(m_u+m_d)=-2\langle \bar{q}q\rangle/f_{\pi}$ from the Gell-Mann-Oakes-Renner relation.

 We fit the free parameters to be
 $C_{1}=-0.0002\,T^2\,\rm{GeV}^6$, $C_{2}=-0.0002\,T^2\,\rm{GeV}^5$, $C_{3}=0.0006\,T^2\,\rm{GeV}^6$,  $C_{4}=-0.00027\,T^2\,\rm{GeV}^5$, $C_{5}=0.00009\,T^2\,\rm{GeV}^5$, $C_{6}=0$, $C_{7}=0.00014\,T^2\,\rm{GeV}^4$, $C_{8}=-0.0002\,T^2\,\rm{GeV}^5$ and $C_{9}=-0.0003\,T^2\,\rm{GeV}^6$ in calculations. And obtain the Borel platforms
 $T^2_{X_0\eta_c\pi}=(4.3-5.3)\,\rm{GeV}^2$,
 $T^2_{X_0\chi_{c1}\pi}=(2.7-3.7)\,\rm{GeV}^2$,
 $T^2_{X_0J/\psi\rho}=(3.1-4.1)\,\rm{GeV}^2$,
 $T^2_{X_1 J/\psi\pi}=(2.3-3.3)\,{\rm GeV}^2$,
 $T^2_{X_1 \eta_c\rho}=(2.2-3.2)\,{\rm GeV}^2$,
 $T^2_{X_1 h_c\pi}=(4.5-5.5)\,{\rm GeV}^2$,
 $T^2_{X_2 \eta_c \pi}=(2.4-3.4)\,{\rm GeV}^2$,
 $T^2_{X_2 \chi_{c1}\pi}=(2.5-3.5)\,{\rm GeV}^2$ and
 $T^2_{X_2 J/\psi\rho}=(2.1-3.1)\,{\rm GeV}^2$,
   where we add the subscripts $X_0\eta_c\pi$, $X_0\chi_{c1}\pi$,
 $X_0J/\psi\rho$, $X_1 J/\psi\pi$, $X_1 \eta_c\rho$, $X_1 h_c\pi$, $X_2\eta_c \pi$, $X_2\chi_{c1}\pi$ and
 $X_2J/\psi\rho$ to denote the corresponding channels.

\begin{figure}
\centering
\includegraphics[totalheight=6cm,width=9cm]{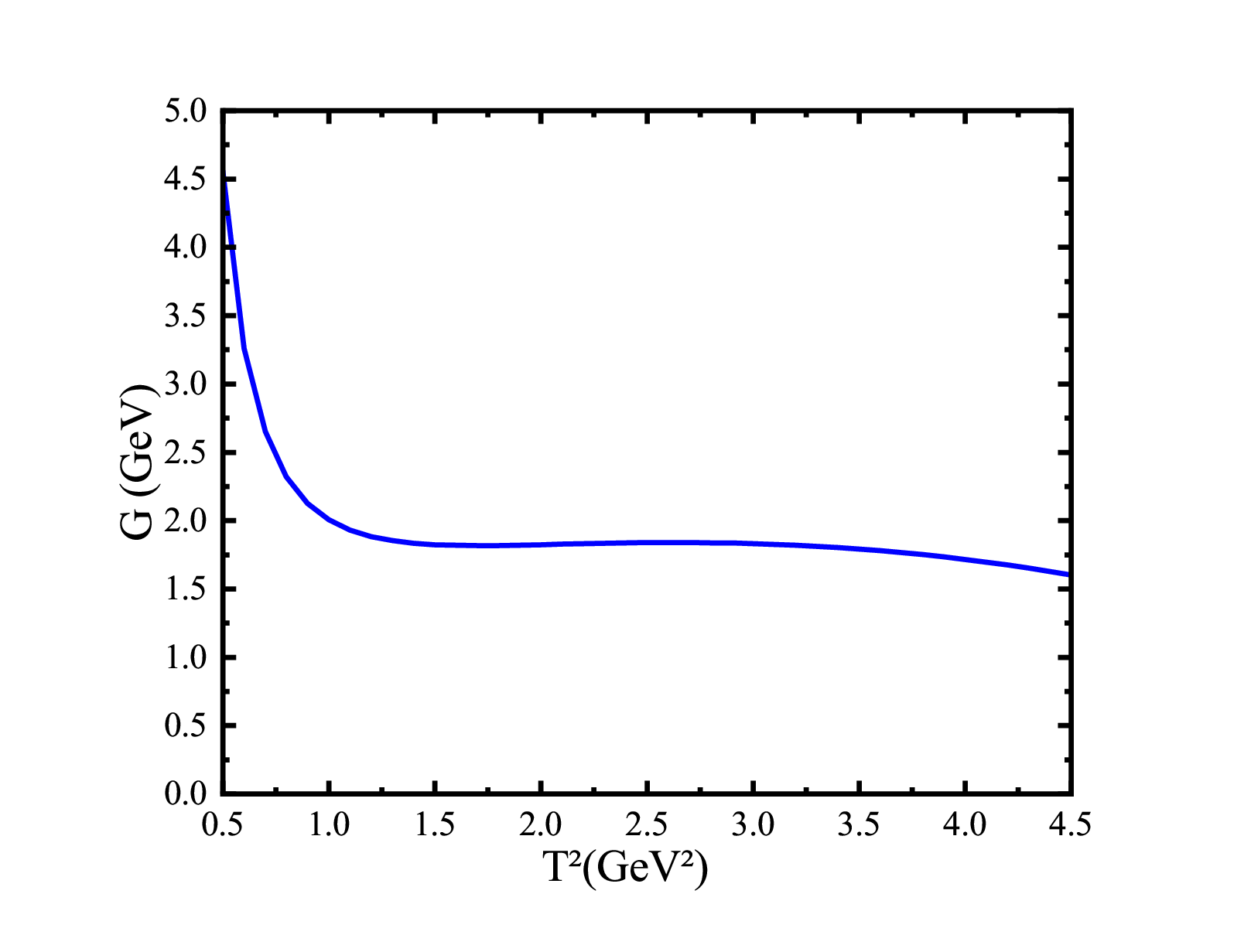}
\caption{The  hadronic coupling constant $G_{X_2 J/\psi\rho}$ with variation of the  Borel  parameter  $T^2$.}\label{hadron-coupling-fig}
\end{figure}

We obtain uniform flat platforms  $T^2_{max}-T^2_{min}=1\,\rm{GeV}^2$ for all the channels, where the max and min denote the maximum and minimum, respectively, just like what have been done in our previous works. For example, in Fig.\ref{hadron-coupling-fig},
we plot the hadronic coupling constant $G_{X_2 J/\psi\rho}$ with variation of the Borel parameter, there appear rather  flat platform indeed, it is reliable to extract the hadronic  coupling constant.

We estimate the uncertainties in the same way as our usually done. For example, the  uncertainty of an input parameter $\xi$, $\xi= \bar{\xi} +\delta \xi$, results in the uncertainties $\lambda_{X_0}f_{J/\psi}f_{\rho}G_{X_0J/\psi \rho} = \bar{\lambda}_{X_0}\bar{f}_{J/\psi}\bar{f}_{\rho}\bar{G}_{X_0J/\psi \rho}
+\delta\,\lambda_{X_0}f_{J/\psi}f_{\rho}G_{X_0J/\psi \rho}$ and $C_{2} = \bar{C}_{2}+\delta C_{2}$,
where
\begin{eqnarray}\label{Uncertainty-4}
	\delta\,\lambda_{X_0}f_{J/\psi}f_{\rho}G_{X_0J/\psi \rho} &=&\bar{\lambda}_{X_0}\bar{f}_{J/\psi} \bar{f}_{\rho} \bar{G}_{X_0J/\psi\rho}\left( \frac{\delta f_{J/\psi}}{\bar{f}_{J/\psi}} +\frac{\delta f_{\rho}}{\bar{f}_{\rho}} +\frac{\delta \lambda_{X_0}}{\bar{\lambda}_{X_0}}+\frac{\delta G_{X_0J/\psi\rho}}{\bar{G}_{X_0J/\psi\rho}}\right)\, ,
\end{eqnarray}
 we can set $\delta C_{2}=0$ and $ \frac{\delta f_{J/\psi}}{\bar{f}_{J/\psi}} = \frac{\delta \lambda_{X_0}}{\bar{\lambda}_{X_0}}=\frac{\delta G_{X_0J/\psi\rho}}{\bar{G}_{XJ/\psi\rho}}$ approximately.
 Finally, we obtain the values of the hadronic coupling constants,
 \begin{eqnarray} \label{HCC-values}
 G_{X_0\eta_c \pi} &=&0.63^{+0.07}_{-0.04}\,\rm{GeV}\, , \nonumber\\
 G_{X_0\chi_{c1}\pi} &=&1.20^{+0.01}_{-0.01}\, , \nonumber\\
 G_{X_0J/\psi\rho} &=&1.79^{+0.22}_{-0.16}\,\rm{GeV}\, , \nonumber\\
 G_{X_1 J/\psi\pi}&=& 6.80^{+0.40}_{-0.38}\,{\rm GeV},\nonumber\\
 G_{X_1 \eta_c\rho}&=& 1.55^{+0.11}_{-0.09}\,{\rm GeV},\nonumber\\
 G_{X_1 h_c\pi}&=& 0.10\,{\rm GeV}^{-1},\nonumber\\
 G_{X_2 \eta_c \pi}&=& 1.49^{+0.02}_{-0.02}\,{\rm GeV}^{-1}\,,\nonumber\\
 G_{X_2\chi_{c1}\pi}&=& 1.20^{+0.02}_{-0.02}\, , \nonumber\\
 G_{X_2 J/\psi\rho}&=& 1.84^{+0.17}_{-0.13}\,{\rm GeV}	\,	,
 \end{eqnarray}
where we have taken the absolute values.

For the molecule masses,  we take the values $m_{X_0}=4.02\,\rm{GeV}$, $m_{X_1}=4.02\,\rm{GeV}$ and $m_{X_2}=4.02\,\rm{GeV}$ from the QCD sum rules \cite{WZG-mole-IJMPA}.
Then we obtain the partial decay widths directly,
\begin{eqnarray} \label{X0-decay}
\Gamma(X_0\to \eta_c\pi)&=& 0.87^{+0.21}_{-0.10} \,\rm{MeV}\, , \nonumber\\
\Gamma(X_0\to \chi_{c1}\pi)&=& 0.45^{+0.01}_{-0.01} \,\rm{MeV}\, , \nonumber\\	
\Gamma(X_0\to J/\psi\rho)&=& 12.33^{+3.22}_{-2.11} \,\rm{MeV}\, ,
\end{eqnarray}
\begin{eqnarray}\label{X1-decay}
\Gamma(X_1 \rightarrow J/\psi\pi)&=& 94.10^{+11.40}_{-10.22}\, {\rm MeV}\, ,\nonumber\\
\Gamma(X_1 \rightarrow \eta_c\rho)&=& 4.22^{+0.62}_{-0.48}\, {\rm MeV}\, ,\nonumber\\	
\Gamma(X_1 \rightarrow h_c\pi)&=& 0.0236\, {\rm MeV}\, ,
\end{eqnarray}
\begin{eqnarray}\label{X2-decay}
\Gamma(X_2 \rightarrow \eta_c \pi)&=& 0.42^{+0.01}_{-0.01}\, {\rm MeV}\, ,\nonumber\\
\Gamma(X_2 \rightarrow \chi_{c1}\pi)&=& 0.12^{+0.01}_{-0.01}\, {\rm MeV}\, ,\nonumber\\
\Gamma(X_2 \rightarrow J/\psi\rho)&=& 4.10^{+0.79}_{-0.56}\, {\rm MeV}\, .
\end{eqnarray}

Then we obtain the total decay  widths approximately,
\begin{eqnarray}
\Gamma_{X_0}&=&13.65^{+3.44}_{-2.22}\,\rm{MeV}\, ,  \nonumber\\
\Gamma_{X_1}&=& 98.34^{+12.02}_{-10.70}\, {\rm MeV}\, , \nonumber\\
\Gamma_{X_2}&=& 4.64^{+0.81}_{-0.58}\, {\rm MeV}\, .
\end{eqnarray}

The predicted width  $\Gamma_{X_2}=4.64^{+0.81}_{-0.58}\,\rm{MeV}$ is  in very good agreement with the experimental data of the   width $(4 \pm 11 \pm 6)~\mathrm{MeV}$ of the $X(4014)$  from the Belle collaboration \cite{Cai:2024glz}. Both the predicted mass and width support assigning  the $X_2(4014)$  as the  $D^*\bar{D}^*$ molecular state with the $J^{PC}=2^{++}$. As for the  predictions  $\Gamma_{X_0}=13.65^{+3.44}_{-2.22}\,\rm{MeV}$ and $\Gamma_{X_1}=98.34^{+12.02}_{-10.70}\,\rm{MeV}$, we can  confront them to the experimental data in the future.

If only the mass is concerned, the $Z_c(4020)$ can be assigned as the diquark-antidiquark type tetraquark state or the $D^*\bar{D}^*$ type molecular state with the $J^{PC}=1^{+-}$ \cite{WZG-mole-IJMPA,WZG-HC-spectrum-PRD}. In the scenario of tetraquark state, we obtain the prediction $\Gamma_{Z_{c}} =29.57\pm2.30\,\, ({\rm or}\, \pm 9.20)\,\,\rm{MeV}$ \cite{WZG-Zcs4123-decay}, which is compatible with the upper  bound of the experimental data  $\Gamma=(24.8\pm5.6\pm7.7)\,\rm{MeV}$ \cite{BES1308},  $(23.0\pm 6.0\pm 1.0)\,\rm{MeV}$ \cite{BES1507}, $(7.9\pm 2.7\pm 2.6)\,\rm{MeV}$ \cite{BES1309} from the BESIII collaboration. While in the molecule scenario, the predicted width  $\Gamma_{X_1}=98.34^{+12.02}_{-10.70}\,\rm{MeV}$ for the $D^*\bar{D}^*$ molecular state with the $J^{PC}=1^{+-}$ is too large compared to the experimental data for the $Z_c(4020)$. The two scenarios lead to quite different predictions for the total widths, and the QCD sum rules favor assigning the $Z_c(4020)$ as the diquark-antidiquark type tetraquark state. Unfortunately,  in our unique scheme, the decay widths of the tetraquark states with the  $J^{PC}=0^{++}$ and $2^{++}$ have not been studied yet.

We can easily determine the relative branching ratios of the tetraquark molecular states from their partial decay widths,
\begin{eqnarray}
	\Gamma\left(X_0 \to \eta_c \pi:\chi_{c1}\pi:J/\psi\rho\right) &=& 0.0710:0.0365:1.00\, ,\nonumber\\
	\Gamma\left(X_1 \to \eta_c\rho:h_c\pi:J/\psi\pi\right) &=& 0.0448:0.0003:1.00\, ,\nonumber\\
	\Gamma\left(X_2 \to \eta_c \pi:\chi_{c1}\pi:J/\psi\rho\right) &=& 0.1024:0.0293:1.00\, .
\end{eqnarray}
Due to the particular quark structures, the dominant decay modes are $X_{0/2} \to J/\psi \rho$ and $X_1 \to J/\psi\pi$, which are consistent with the observation of the $X_2(4014)$ in the $\gamma \psi(2S)$ mass spectrum with the vector meson dominance $X_2 \to \gamma^* J/\psi \to \rho J/\psi$,   we can search for those molecular states in those typical decay channels.

\section{Conclusion}
In this work, we study the hadronic coupling constants in the two-body strong decays of the tetraquark molecular states with the $J^{PC}=0^{++}$, $1^{+-}$ and $2^{++}$ via the three-point correlation functions.
We carry out the operator product expansion considering  the quark condensate, gluon condensate and quark-gluon mixed condensate
 to obtain the QCD spectral representations, then match the QCD side with the hadron side  according to rigorous quark-hadron duality.
We obtain the hadronic coupling constants and partial decay widths therefore total  widths
of the tetraquark molecular states with the $J^{PC}=0^{++}$, $1^{+-}$ and $2^{++}$, respectively. Our predictions indicate assigning the $X_2(4014)$ as the $D^*\bar{D}^*$ tetraquark molecular states with the $J^{PC}=2^{++}$ is reasonable, while other predictions  play an interesting role in diagnosing the $X$, $Y$ and $Z$ states. Moreover, the $X_2(4014)$ still needs independent confirmation by other experiments.
	
\section*{Appendix}
The analytical expressions of the other QCD sum rules,
\begin{eqnarray}\label{QCDSR-X0-chi_{c1}-pi}	
	&&\frac{\lambda_{X_0\chi_{c1}\pi} G_{X_0\chi_{c1}\pi}}{ {m}^2_{X_0}-m^2_{\chi_{c1}}}\left[\exp\left(-\frac{m^2_{\chi_{c1}}}{T^2}\right) -\exp\left(-\frac{{m}^2_{X_0}}{T^2}\right) \right] \exp\left(-\frac{m^2_{\pi}}{T^2}\right)+C_{2} \exp\left(-\frac{m^2_{\chi_{c1}}}{T^2}-\frac{m^2_{\pi}}{T^2}\right)  \nonumber\\
	&=&\frac{\langle\bar{q}q\rangle}{8\pi^2}\int_{4m_c^2}^{s^0_{\chi_{c1}}} ds
	 \left(s-4m^2_c\right)\sqrt{1-\frac{4m_c^2}{s}}\exp\left(-\frac{s}{T^2}\right)\nonumber\\
	&&-\frac{\langle\bar{q}g_{s}\sigma Gq\rangle}{24\pi^2}\int_{4m_c^2}^{s^0_{\chi_{c1}}}ds\frac{ \left(s-6m_c^2\right)}{ \sqrt{s\left(s-4m^2_c\right)}}\exp\left(-\frac{s}{T^2}\right)  \, ,	
\end{eqnarray}

\begin{eqnarray}\label{QCDSR-X0-J/psi-rho}	
	&&\frac{\lambda_{X_0J/\psi\rho} G_{X_0J/\psi\rho}}{ {m}^2_{X_0}-m^2_{J/\psi}}\left[\exp\left(-\frac{m^2_{J/\psi}}{T^2}\right) -\exp\left(-\frac{{m}^2_{X_0}}{T^2}\right) \right] \exp\left(-\frac{m^2_{\rho}}{T^2}\right)+C_{3} \exp\left(-\frac{m^2_{J/\psi}}{T^2}-\frac{m^2_{\rho}}{T^2}\right)  \nonumber\\
	&=&\frac{3}{128\pi^4}  \int^{s^0_{J/\psi}}_{4m^2_c} ds \int^{s^0_{\rho}}_{0} du su\sqrt{1-\frac{4m^2_c}{s}}\exp\left(-\frac{s+u}{T^2}\right) \nonumber\\
	&&+\frac{m^4_c }{16\pi^2}\langle\frac{\alpha_{s}GG}{\pi}\rangle \int^{s^0_{J/\psi}}_{4m^2_c} ds \int^{s^0_{\rho}}_{0} du \frac{su  \left(2m^2_c-s\right)} {\sqrt{s\left(s-4m^2_c\right)}^5} \exp\left(-\frac{s+u}{T^2}\right)  \nonumber\\
	&&+\frac{m^4_c }{96\pi^2}\langle\frac{\alpha_{s}GG}{\pi}\rangle \int^{s^0_{J/\psi}}_{4m^2_c} ds \int^{s^0_{\rho}}_{0} du \frac{u} {\sqrt{s\left(s-4m^2_c\right)}^3} \exp\left(-\frac{s+u}{T^2}\right)  \nonumber\\
	&&-\frac{1}{192\pi^2}\langle\frac{\alpha_{s}GG}{\pi}\rangle \int^{s^0_{J/\psi}}_{4m^2_c} ds \int^{s^0_{\rho}}_{0} du \frac{m^2_c+s} {\sqrt{s\left(s-4m^2_c\right)}} \exp\left(-\frac{s+u}{T^2}\right)  \, ,
\end{eqnarray}

\begin{eqnarray}\label{QCDSR-X1-J/psi-pi}	
	&&\frac{\lambda_{X_1J/\psi\pi} G_{X_1J/\psi\pi}}{ {m}^2_{X_1}-m^2_{J/\psi}}\left[\exp\left(-\frac{m^2_{J/\psi}}{T^2}\right) -\exp\left(-\frac{{m}^2_{X_1}}{T^2}\right) \right] \exp\left(-\frac{m^2_{\pi}}{T^2}\right)+C_{4} \exp\left(-\frac{m^2_{J/\psi}}{T^2}-\frac{m^2_{\pi}}{T^2}\right)  \nonumber\\
	&=&-\frac{m_c}{16\sqrt{2}\pi^4}  \int^{s^0_{J/\psi}}_{4m^2_c} ds \int^{s^0_{\pi}}_{0} du u\sqrt{1-\frac{4m^2_c}{s}}\exp\left(-\frac{s+u}{T^2}\right) \nonumber\\
	&&-\frac{m_c }{48\sqrt{2}\pi^2}\langle\frac{\alpha_{s}GG}{\pi}\rangle \int^{s^0_{J/\psi}}_{4m^2_c} ds \int^{s^0_{\pi}}_{0} du \frac{su  \left(20m^4_c-10sm^2_c+s^2\right)} {\sqrt{s\left(s-4m^2_c\right)}^5} \exp\left(-\frac{s+u}{T^2}\right)  \nonumber\\
	&&+\frac{m_c }{144\sqrt{2}\pi^2}\langle\frac{\alpha_{s}GG}{\pi}\rangle \int^{s^0_{J/\psi}}_{4m^2_c} ds \int^{s^0_{\pi}}_{0} du \frac{u\left(s-2m^2_c\right)} {s\left(4m^2_c-s\right)\sqrt{s\left(s-4m^2_c\right)}} \exp\left(-\frac{s+u}{T^2}\right)  \nonumber\\
	&&+\frac{m_c}{144\sqrt{2}\pi^2}\langle\frac{\alpha_{s}GG}{\pi}\rangle \int^{s^0_{J/\psi}}_{4m^2_c} ds \int^{s^0_{\pi}}_{0} du \frac{3m^2_c+s} {s\sqrt{s\left(s-4m^2_c\right)}} \exp\left(-\frac{s+u}{T^2}\right) \nonumber \\
	&&+\frac{\langle\bar{q}q\rangle}{12\sqrt{2}\pi^2}\int_{4m_c^2}^{s^0_{J/\psi}} ds
	 \left(2m^2_c+s\right)\sqrt{1-\frac{4m_c^2}{s}}\exp\left(-\frac{s}{T^2}\right)\nonumber\\
	&&+\frac{\langle\bar{q}g_{s}\sigma Gq\rangle}{32\sqrt{2}\pi^2T^2}\int_{4m_c^2}^{s^0_{J/\psi}}ds\left(2m^2_c+s\right)\sqrt{1-\frac{4m_c^2}{s}}
\exp\left(-\frac{s}{T^2}\right)\nonumber\\
	&&-\frac{\langle\bar{q}g_{s}\sigma Gq\rangle}{64\sqrt{2}\pi^2}\int_{4m_c^2}^{s^0_{J/\psi}}ds\frac{4m^2_c+s} {s} \sqrt{1-\frac{4m_c^2}{s}}\exp\left(-\frac{s}{T^2}\right)\nonumber\\
	&&-\frac{5m^2_c\langle\bar{q}g_{s}\sigma Gq\rangle}{24\sqrt{2}\pi^2}\int_{4m_c^2}^{s^0_{J/\psi}}ds\frac{1} {\sqrt{s\left(s-4m_c^2\right)}} \exp\left(-\frac{s}{T^2}\right)  \, ,
\end{eqnarray}

\begin{eqnarray}\label{QCDSR-X1-eta_c-rho}	
	&&\frac{\lambda_{X_1\eta_c\rho} G_{X_1\eta_c\rho}}{ {m}^2_{X_1}-m^2_{\eta_c}}\left[\exp\left(-\frac{m^2_{\eta_c}}{T^2}\right) -\exp\left(-\frac{{m}^2_{X_1}}{T^2}\right) \right] \exp\left(-\frac{m^2_{\rho}}{T^2}\right)+C_{5} \exp\left(-\frac{m^2_{\eta_c}}{T^2}-\frac{m^2_{\rho}}{T^2}\right)  \nonumber\\
	&=&\frac{m_c}{64\sqrt{2}\pi^4}  \int^{s^0_{\eta_c}}_{4m^2_c} ds \int^{s^0_{\rho}}_{0} du \frac{u^2+2su}{s}\sqrt{1-\frac{4m^2_c}{s}}\exp\left(-\frac{s+u}{T^2}\right) \nonumber\\
	&&+\frac{m_c }{192\sqrt{2}\pi^2}\langle\frac{\alpha_{s}GG}{\pi}\rangle \int^{s^0_{\eta_c}}_{4m^2_c} ds \int^{s^0_{\rho}}_{0} du \frac{u\left(2s+u\right)\left(20m^4_c-10sm^2_c+s^2\right)} {\sqrt{s\left(s-4m^2_c\right)}^5} \exp\left(-\frac{s+u}{T^2}\right)  \nonumber\\
	&&-\frac{m_c }{192\sqrt{2}\pi^2}\langle\frac{\alpha_{s}GG}{\pi}\rangle \int^{s^0_{\eta_c}}_{4m^2_c} ds \int^{s^0_{\rho}}_{0} du \frac{u\left(2s+u\right)\left(s-2m^2_c\right)} {s\sqrt{s\left(s-4m^2_c\right)}^3} \exp\left(-\frac{s+u}{T^2}\right)  \nonumber\\
	&&+\frac{m_c }{576\sqrt{2}\pi^2}\langle\frac{\alpha_{s}GG}{\pi}\rangle \int^{s^0_{\eta_c}}_{4m^2_c} ds \int^{s^0_{\rho}}_{0} du \frac{2s+u} {s\sqrt{s\left(s-4m^2_c\right)}} \exp\left(-\frac{s+u}{T^2}\right)  \nonumber\\
	&&-\frac{11\langle\bar{q}g_{s}\sigma Gq\rangle}{144\sqrt{2}\pi^2}\int_{4m_c^2}^{s^0_{\eta_c}}ds \sqrt{1-\frac{4m_c^2}{s}}\exp\left(-\frac{s}{T^2}\right)  \, ,
\end{eqnarray}

\begin{eqnarray}\label{QCDSR-X1-h_c-pi}	
	&&\frac{\lambda_{X_1h_c\pi} G_{X_1h_c\pi}}{ {m}^2_{X_1}-m^2_{h_c}}\left[\exp\left(-\frac{m^2_{h_c}}{T^2}\right) -\exp\left(-\frac{{m}^2_{X_1}}{T^2}\right) \right] \exp\left(-\frac{m^2_{\pi}}{T^2}\right)+C_{6} \exp\left(-\frac{m^2_{h_c}}{T^2}-\frac{m^2_{\pi}}{T^2}\right)  \nonumber\\
	&=&\frac{1}{384\sqrt{2}\pi^4}  \int^{s^0_{h_c}}_{4m^2_c} ds \int^{s^0_{\pi}}_{0} du \frac{u\left(4m^2_c-s\right)}{s^2}\sqrt{1-\frac{4m^2_c}{s}}\exp\left(-\frac{s+u}{T^2}\right) \nonumber\\
	&&+\frac{m^2_c }{576\sqrt{2}\pi^2}\langle\frac{\alpha_{s}GG}{\pi}\rangle \int^{s^0_{h_c}}_{4m^2_c} ds \int^{s^0_{\pi}}_{0} du \frac{u\left(2m^2_c+s\right)} {s^3\left(4m^2_c-s\right)\sqrt{s\left(s-4m^2_c\right)}} \exp\left(-\frac{s+u}{T^2}\right)  \nonumber\\
	&&+\frac{m^2_c}{1728\sqrt{2}\pi^2}\langle\frac{\alpha_{s}GG}{\pi}\rangle \int^{s^0_{h_c}}_{4m^2_c} ds \int^{s^0_{\pi}}_{0} du \frac{u} {s\sqrt{s\left(s-4m^2_c\right)}^3} \exp\left(-\frac{s+u}{T^2}\right)  \nonumber\\
	&&+\frac{1}{3456\sqrt{2}\pi^2}\langle\frac{\alpha_{s}GG}{\pi}\rangle \int^{s^0_{h_c}}_{4m^2_c} ds \int^{s^0_{\pi}}_{0} du \frac{12m^2_c-5s} {s^2\sqrt{s\left(s-4m^2_c\right)}} \exp\left(-\frac{s+u}{T^2}\right)\nonumber\\
	&&+\frac{25m^3_c \langle\bar{q}g_{s}\sigma Gq\rangle}{108\sqrt{2}\pi^2}\int_{4m_c^2}^{s^0_{h_c}}ds\frac{ 1}{s^2 \sqrt{s\left(s-4m^2_c\right)}}\exp\left(-\frac{s}{T^2}\right)\nonumber\\	
	&&-\frac{5m_c \langle\bar{q}g_{s}\sigma Gq\rangle}{108\sqrt{2}\pi^2}\int_{4m_c^2}^{s^0_{h_c}}ds\frac{1}{s \sqrt{s\left(s-4m^2_c\right)}}\exp\left(-\frac{s}{T^2}\right)  \, ,
\end{eqnarray}

\begin{eqnarray}\label{QCDSR-X2-eta-pi}	
	&&\frac{\lambda_{X_2\eta_c\pi} G_{X_2\eta_c\pi}}{ {m}^2_{X_2}-m^2_{\eta_c}}\left[\exp\left(-\frac{m^2_{\eta_c}}{T^2}\right) -\exp\left(-\frac{{m}^2_{X_2}}{T^2}\right) \right] \exp\left(-\frac{m^2_{\pi}}{T^2}\right)+C_{7} \exp\left(-\frac{m^2_{\eta_c}}{T^2}-\frac{m^2_{\pi}}{T^2}\right)  \nonumber\\
	&=&-\frac{m_c \langle\bar{q}q\rangle}{2\sqrt{2}\pi^2}\int_{4m_c^2}^{s^0_{\eta_c}} ds
	\sqrt{1-\frac{4m_c^2}{s}}\exp\left(-\frac{s}{T^2}\right)\nonumber\\
	&&+\frac{m_c \langle\bar{q}g_{s}\sigma Gq\rangle}{8\sqrt{2}\pi^2T^2}\int_{4m_c^2}^{s^0_{\eta_c}}ds\sqrt{1-\frac{4m_c^2}{s}}
\exp\left(-\frac{s}{T^2}\right)\nonumber\\
	&&-\frac{5m_c \langle\bar{q}g_{s}\sigma Gq\rangle}{24\sqrt{2}\pi^2}\int_{4m_c^2}^{s^0_{\eta_c}}ds
\frac{ 1}{ \sqrt{s\left(s-4m^2_c\right)}}\exp\left(-\frac{s}{T^2}\right)\nonumber\\
	&&-\frac{5m_c \langle\bar{q}g_{s}\sigma Gq\rangle}{16\sqrt{2}\pi^2T^2}\int_{4m_c^2}^{s^0_{\eta_c}}ds\frac{ \sqrt{s\left(s-4m^2_c\right)}}{s}\exp\left(-\frac{s}{T^2}\right)  \, ,
\end{eqnarray}

\begin{eqnarray}\label{QCDSR-X2-chi_{c1}-pi}	
	&&\frac{\lambda_{X_2\chi_{c1}\pi} G_{X_2\chi_{c1}\pi}}{ {m}^2_{X_2}-m^2_{\chi_{c1}}}\left[\exp\left(-\frac{m^2_{\chi_{c1}}}{T^2}\right) -\exp\left(-\frac{{m}^2_{X_2}}{T^2}\right) \right] \exp\left(-\frac{m^2_{\pi}}{T^2}\right)+C_{8} \exp\left(-\frac{m^2_{\chi_{c1}}}{T^2}-\frac{m^2_{\pi}}{T^2}\right)  \nonumber\\
	 &=&-\frac{\langle\bar{q}q\rangle}{8\sqrt{2}\pi^2}\int_{4m_c^2}^{s^0_{\chi_{c1}}} ds
	 \left(4m^2_c-s\right)\sqrt{1-\frac{4m_c^2}{s}}\exp\left(-\frac{s}{T^2}\right)\nonumber\\
	&&-\frac{5\langle\bar{q}g_{s}\sigma Gq\rangle}{48\sqrt{2}\pi^2}\int_{4m_c^2}^{s^0_{\chi_{c1}}}ds\frac{ \left(6m_c^2-s\right)}{ \sqrt{s\left(s-4m^2_c\right)}}\exp\left(-\frac{s}{T^2}\right)\nonumber\\	
	&&-\frac{7\langle\bar{q}g_{s}\sigma Gq\rangle}{64\sqrt{2}\pi^2T^2}\int_{4m_c^2}^{s^0_{\chi_{c1}}}ds\left(4m^2_c-s\right)\sqrt{1-\frac{4m_c^2}{s}}
 \exp\left(-\frac{s}{T^2}\right)  \, ,
\end{eqnarray}

\begin{eqnarray}\label{QCDSR-X2-J/psi-rho}	
	&&\frac{\lambda_{X_2J/\psi\rho} G_{X_2J/\psi\rho}}{ {m}^2_{X_2}-m^2_{J/\psi}}\left[\exp\left(-\frac{m^2_{J/\psi}}{T^2}\right) -\exp\left(-\frac{{m}^2_{X_2}}{T^2}\right) \right] \exp\left(-\frac{m^2_{\rho}}{T^2}\right)+C_{9} \exp\left(-\frac{m^2_{J/\psi}}{T^2}-\frac{m^2_{\rho}}{T^2}\right)  \nonumber\\
	&=&-\frac{1}{96\sqrt{2}\pi^4}  \int^{s^0_{J/\psi}}_{4m^2_c} ds \int^{s^0_{\rho}}_{0} du u\left(s+2m^2_c\right)\sqrt{1-\frac{4m^2_c}{s}}\exp\left(-\frac{s+u}{T^2}\right) \nonumber\\
	&&-\frac{m^2_c }{288\sqrt{2}\pi^2}\langle\frac{\alpha_{s}GG}{\pi}\rangle \int^{s^0_{J/\psi}}_{4m^2_c} ds \int^{s^0_{\rho}}_{0} du \frac{su\left(48m^4_c-22sm^2_c+s^2\right)} {\sqrt{s\left(s-4m^2_c\right)}^5} \exp\left(-\frac{s+u}{T^2}\right)  \nonumber\\
	&&+\frac{m^2_c }{288\sqrt{2}\pi^2}\langle\frac{\alpha_{s}GG}{\pi}\rangle \int^{s^0_{J/\psi}}_{4m^2_c} ds \int^{s^0_{\rho}}_{0} du \frac{u\left(6m^2_c-s\right)} {s\left(4m^2_c-s\right)\sqrt{s\left(s-4m^2_c\right)}} \exp\left(-\frac{s+u}{T^2}\right)  \, ,
\end{eqnarray}
where we introduce the notations,
\begin{eqnarray}
	\lambda_{X_0\chi_{c1}\pi}&=&\frac{\lambda_{X_0}f_{\chi_{c1}} m_{\chi_{c1}} f_{\pi} m_{\pi}^2}{m_u+m_d}\, , \nonumber \\
	\lambda_{X_0J/\psi\rho}&=&\lambda_{X_0} f_{J/\psi} m_{J/\psi} f_{\rho} m_{\rho}\, ,
\end{eqnarray}
\begin{eqnarray}
	\lambda_{X_1J/\psi\pi}&=&\frac{\lambda_{X_1} f_{J/\psi} m_{J/\psi} f_{\pi} m^2_{\pi}}{\left(m_u+m_d\right)m_{X_1}}\, , \nonumber\\
	\lambda_{X_1\eta_c\rho}&=&\frac{\lambda_{X_1} f_{\eta_c} m^2_{\eta_c} f_{\rho} m_{\rho}}{2m_cm_{X_1}}\, , \nonumber\\
	\lambda_{X_1h_c\pi}&=&\frac{2\lambda_{X_1}\lambda_{h_c} f_{\pi} m^2_{\pi}}{9\left(m_u+m_d\right)m_{X_1}}\, ,
\end{eqnarray}
\begin{eqnarray}
	\lambda_{X_2\eta_c\pi}&=&\frac{\lambda_{X_2} f_{\eta_c} m^2_{\eta_c} f_{\pi} m^2_{\pi} (m_{X_2}^2-m_{\eta_c}^2)}{2m_c\left(m_u+m_d\right)m_{X_2}^2}\, ,\nonumber \\
	\lambda_{X_2\chi_{c1}\pi}&=&\frac{\lambda_{X_2} f_{\chi_{c1}} m_{\chi_{c1}}f_{\pi} m^2_{\pi}}{m_u+m_d}\, ,\nonumber \\
	\lambda_{X_2J/\psi\rho}&=&\frac{\lambda_{X_2} f_{J/\psi} m_{J/\psi}f_{\rho} m_{\rho}}{2} \, .
\end{eqnarray}
	
\section*{Acknowledgements}
This  work is supported by National Natural Science Foundation, Grant Number 12175068.

\end{document}